\documentclass[aps]{revtex4}

\newcommand{\be}{\begin{equation}}
\newcommand{\ee}{\end{equation}}

\newcommand{\bd}{\begin{displaymath}}
\newcommand{\ed}{\end{displaymath}}
\newcommand{\bea}{\begin{eqnarray}}
\newcommand{\eea}{\end{eqnarray}}

\begin{document}

\title{Contribution to understanding the mathematical structure of quantum mechanics}

\author{L. Sk\'ala$^{1,2}$\footnote{Corresponding author. 
E-mail: skala@karlov.mff.cuni.cz}
 and V. Kapsa$^1$}
\affiliation
{$^1$Charles University, Faculty of Mathematics and Physics, 
Ke Karlovu 3, 121 16 Prague 2, Czech Republic} 
\affiliation
{$^2$University of Waterloo, Department of Applied Mathematics,\\
Waterloo, Ontario N2L 3G1, Canada}

\begin{abstract}
Probabilistic description of results of measurements and its consequences for 
understanding quantum mechanics are discussed.
It is shown that the basic mathematical structure of quantum mechanics
like the probability amplitudes, Born rule, 
commutation and uncertainty relations, probability
density current, momentum operator, rules for including the 
scalar and vector potentials and antiparticles can be obtained from
the probabilistic description of results of measurement of the
space coordinates and time.
Equations of motion of quantum mechanics, the Klein-Gordon equation,
Schr\"odinger equation and Dirac equation are obtained from 
the requirement of the relativistic invariance of the space-time
Fisher information.
The limit case of the $\delta$-like probability densities leads to the
Hamilton-Jacobi equation of classical mechanics.
Many particle systems 
and the postulates of quantum mechanics are also discussed.
\end{abstract}

\maketitle 

\noindent
keywords: probability theory, quantum mechanics, classical mechanics

\section{Introduction}

\label{introduction}

Quantum mechanics is one of the most completely tested physical
theories (see e.g. \cite{Scully1,Zeilinger2,Zeilinger4,Bertlmann}).
At the same time, the standard approach to introducing quantum mechanics based 
on the sometimes contra intuitive postulates does not clarify the roots
of quantum mechanics and the exact physical meaning of the postulates
and their interpretation is the subject of continuing discussion 
(see e.g. 
\cite{Birkhoff,Feynman,Bohm1,Bohm2,Bohm3,Everett,Wheeler,Ludwig,Lande,Cramer,Fivel,Bohr,Brukner,Fuchs,Lamb,Hardy,Summhammer,Parwani,Frieden,Roy,Laloe}).
It is not satisfactory and, in our opinion, it
is necessary to concentrate on a more direct description
of the probabilistic character of measurements and its
consequences. 
Such approach can clarify the most important 
assumptions made in quantum mechanics and contribute 
to understanding quantum mechanics as a probabilistic theory
of certain class of physical phenomena.

In this paper, we do not want to develop a new interpretation
of quantum mechanics.
Rather, we would like to contribute to understanding the standard 
or Copenhagen interpretation by illuminating its basic ideas
by means of the probabilistic description of measurements.  
It is to be noted that the approach used in this paper is 
different from that used usually in physics:
To explain experimental results, one introduces some physical quantities and
equations of motion for these quantities. 
Then, consequences of these equations are investigated,
compared with results of measurements and this procedure is repeated.
The probabilistic character of physical phenomena in the quantum world
is well-known and has to be respected in any attempt to
understanding quantum mechanics.
Therefore, we describe results of measurements in a probabilistic way
and ask what is the mathematical apparatus that can describe this
situation in the most simple manner. 
Using such probabilistic or information theoretical approach, 
the basic mathematical structure of quantum mechanics
except for equations of motion is obtained. 
Equations of motion are found from the requirement of the
relativistic invariance of the theory.
This paper is extended version of the papers \cite{Skala}.

Similar approach based on the principle of extreme physical
information was used also by Frieden who derived the most important
equations of motion in physics (see \cite{Frieden}).
 
In this paper, we do not discuss measurement processes in detail
and assume that measuring apparatuses for measuring
the spatial coordinates and time exist.    
Models based on the probabilistic description of the measured system and
measuring apparatus interacting with the thermodynamic bath can be
found for example in \cite{Allah1,Allah2}, see also \cite{Hay}.

Probably the best approach is to start with measurement
of the spatial coordinates and time.
In this paper, we show that the basic mathematical structure of
quantum mechanics like the probability amplitudes, Born rule, 
commutation and uncertainty relations, probability
density current, kinetic energy, momentum operator or the
rules for including the scalar and vector
potentials and antiparticles can be derived from
the probabilistic description of results of measurement of 
the spatial coordinates and time by means of the probability density
and probability density current (sections \ref{Bornrule}-\ref{scalar}).
Equations of motion of quantum mechanics, the Klein-Gordon equation,
Schr\"odinger equation and Dirac equation are obtained from the requirement of
the relativistic invariance of the generalized space--time Fisher information
(section \ref{equations}).
The limit case of the localized probability densities yields the 
Hamilton-Jacobi equation of classical mechanics (section
\ref{classical}). 
Generalization to many particle systems is performed in
section \ref{many}. 
Postulates of quantum mechanics
are discussed in section \ref{postulates}.

\section{Probability density and the Born rule}

\label{Bornrule}

Physical experiments show that results of measurements 
have very often probabilistic character.
It is related to the well-known experimental conditions of measurements:
The interaction of the measured system with the measuring apparatus
and the rest of the world
cannot be in general neglected, measuring apparatuses are
not described in detail but on the macroscopic level only, 
real physical detectors have limited resolution and efficiency, 
the experimental control of the initial conditions
is limited, etc. 
As a result, resolution of physical experiments is always limited
and the assumption that measurements can be arbitrarily exact
(made for example in classical mechanics) is not valid.
Therefore, results of measurements must be described in
a probabilistic manner (see e.g. \cite{Bohr,Summhammer}). 

General definition of the mean value of a real physical
quantity $A$ can be written either in the continuous 
\be
\label{meanAc}
\langle A\rangle=\int A \rho(A) {\rm d}A 
\ee
or discrete form
\be
\label{meanAd}
\langle A\rangle=\sum_i A_i \rho_i.
\ee
Here, $A$ resp. $A_i$ denote continuous resp. discrete values of 
the quantity $A$ that can be obtained in measurements 
and $\rho(A)$ resp. $\rho_i$ are relative weights
of the corresponding probabilistic distributions.

To be more concrete, we will first discuss measurement of the coordinate $x$.
Results of repeated measurements of the coordinate $x$
can be in physically reasonable cases characterized by the mean values of the moments
\be
\label{meanx}
\langle x^n\rangle=\int x^n\rho({\bf r},t){\rm d}V,\quad n=0,1,2,\ldots,
\ee
where the integration is carried out over the whole space, 
${\rm d}V={\rm d}x{\rm d}y{\rm d}z$ and $\rho({\bf r},t)\geq 0$
is a normalized probability density to obtain
the coordinate $x$ in measurement made at time $t$
\be
\label{cnorm}
\int \rho\, {\rm d}V=\langle x^0\rangle=1.
\ee
This normalization condition is supposed to be valid at all times $t$.
  
First, we perform the integration by parts with respect to the variable $x$ in
Eq. (\ref{meanx}) and get
\be
\label{cpx0}
\left. \frac{x^{n+1}}{n+1}\rho\right|_{x=-\infty}^{\infty}-
\int \frac{x^{n+1}}{n+1}\frac{\partial\rho}{\partial x}{\rm d}V
=\langle x^n\rangle.
\ee
Assuming that the first term in this equation equals zero 
for physically reasonable $\rho$ 
(for example, for the bound states) we get 
\be
\label{cpx1}
\int x^{n+1}\frac{\partial\rho}{\partial x}{\rm d}V=
-(n+1)\langle x^n\rangle,\quad n=0,1,2,\ldots.
\ee
We will show that this simple result
has consequences interesting from the point
of view of the basic structure of quantum mechanics.

The last equation can be rewritten in form of the inner product
\be
\label{cuvx}
(u,v)=-(n+1)\langle x^n\rangle
\ee
defined in the usual way
\be
\label{cscalarx}
(u,v)=\int u^*v\,{\rm d}V.
\ee
Here, the star denotes the complex conjugate and the functions $u$ and $v$
can be taken in general form
\be
\label{cux}
u=x^{n+1}\psi,
\ee
\be
\label{cvx}
v=\frac{1}{\psi^*}\frac{\partial\rho}{\partial x},
\ee
where $\psi=\psi({\bf r},t)$ is an arbitrary complex function.
We note that Eq. (\ref{cuvx}) has the same physical and mathematical
content as Eq. (\ref{cpx1}). 
Our aim is to find conditions for the function $\psi$ that will lead to
the physically most reasonable and mathematically most simple
formulation of the theory. 

Generally valid property of the inner product (\ref{cscalarx}) is the Schwarz
inequality
\be
\label{Schwarzorig}
(u,u)(v,v) \geq |(u,v)|^2.
\ee
Due to Eq. (\ref{cuvx}), the Schwarz inequality yields in our case
\be
\label{Schwarz}
(u,u)(v,v) \geq (n+1)^2\langle x^n\rangle^2,\quad n=0,1,2,\ldots,
\ee
where 
\be
\label{Schwarz1}
(u,u)=\int x^{2n+2}|\psi|^2 {\rm d}V, \quad n=0,1,2,\ldots
\ee
and
\be
\label{vv}
(v,v)=\int\frac{1}{|\psi|^2}\left(\frac{\partial\rho}{\partial x}\right)^2{\rm d}V.
\ee
Till now, $\psi$ could be an arbitrary complex function and the integrals $(u,u)$ and
$(v,v)$ have in general no direct relation to the mean values $\langle x^n\rangle$
characterizing measurement of the coordinate $x$. 
Therefore, inequalities (\ref{Schwarz}) are for general $\psi$ only
a mathematical result without any direct physical meaning. 
 
However, since $\psi$ can be an arbitrary function,
we can require that the integrals $(u,u)$ in Eq. (\ref{Schwarz1}) 
do have a physical meaning and the function $\psi$ obeys the conditions
\be
\label{equal}
\int x^{2n+2}|\psi|^2 {\rm d}V=\int x^{2n+2}\rho\,{\rm d}V,\quad n=0,2,4,\ldots.
\ee
If these conditions are fulfilled, the inequalities (\ref{Schwarz}) contain
the moments $\langle x^n\rangle$ used above
for describing the results of measurements. 
The inner product $(v,v)$ will be discussed below.

Conditions (\ref{equal}) together with the normalization
condition that can be applied to $\psi$
\be
\label{normpsi}
\int \rho\,{\rm d}V=\int |\psi|^2 {\rm d}V=1
\ee 
do not determine the relation between $\rho$ and $|\psi|^2$ uniquely.
They show, however, that the most simple physically reasonable
relation between the probability density $\rho$ and 
probability amplitude $\psi$ has the form of the Born rule
\be
\label{Born}
\rho=|\psi|^2
\ee
or
\be
\label{psi}
\psi=\sqrt{\rho}\, {\rm e}^{is_1},
\ee
where $s_1=s_1({\bf r},t)$ is a real function
that will be discussed in section \ref{probability}.
It follows from the last two equations that
the function $\psi$ can be called the probability amplitude.

Defining the function $s_2=s_2({\bf r},t)$ by
the equation 
\be
\sqrt{\rho}={\rm e}^{-s_2} 
\ee
the probability amplitude $\psi=\psi({\bf r},t)$ 
can be written also in the ``eikonal'' form
\be
\label{psieik}
\psi={\rm e}^{is},
\ee
where $s_1$ and $s_2$ are the real and imaginary parts of $s$
\be
s=s_1+is_2
\ee
and the function $s_2=-(1/2)\ln\rho$ gives the form of the probability
density $\rho$.

We have seen that the integration by parts applied to the definition 
of the mean values (\ref{meanx}) yields Eq. (\ref{cpx1}) that has
interesting physical and mathematical implications.
Its general consequence in form of the inequality
(\ref{Schwarz}) does not contain physically relevant quantities
unless we assume that the function $\psi$ obeys conditions 
(\ref{equal}). 
The most simple solution of these conditions and the normalization
condition (\ref{normpsi}) has the form of the well-known Born rule (\ref{Born}) 
(\cite{Born}, see also \cite{Bohm3,Cramer,Ozawa}).

It is to be noted that our way of writing the probability amplitude
in Eq. (\ref{psieik}) is very similar to the expression used by Bohm
\cite{Bohm1,Bohm2} (see also Madelung \cite{Madelung})
\be
\psi=R\,{\rm e}^{iS/\hbar}.
\ee
To get the same formula as Bohm, we can put $\hbar=1$, $S=s_1$ and 
$R=\exp(-s_2)$.
For this reason, it is not surprising that some parts of the following 
discussion are similar to that performed by Bohm.
The most important differences between our approach
and that of Bohm are summarized in Conclusions.

\section{Commutation relation}

\label{commutation}

Now we return back to the normalization condition
\be
\label{normpsi1}
\int|\psi|^2{\rm d}V=1.
\ee 
Performing the integration by parts in this equation and assuming
that $x|\psi|^2\rightarrow 0$ for $x\rightarrow\pm\infty$
we get 
\be
\label{basic}
\int x\left(\frac{\partial \psi^*}{\partial x}\psi
+\psi^*\frac{\partial \psi}{\partial x}\right)
{\rm d}V=-1.
\ee
Multiplying this equation by $-i$ we obtain the equation
\be
\label{apx2}
\int\left[(x\psi)^*\left(-i\frac{\partial\psi}{\partial x}\right)
-\left(-i\frac{\partial\psi}{\partial x}\right)^*x\psi\right]{\rm d}V=
2i\int x\frac{\partial s_2}{\partial x}{\rm e}^{-2s_2}{\rm d}V=i
\ee
or in the operator form
\be
\label{apx3}
[x, -i(\partial/\partial x)]=i.
\ee

Except for the factor $\hbar$ determining the choice of units, 
this commutation relation
agrees with the commutation relation $[x,\hat{p}_x]=i\hbar$ between the coordinate $x$ and the
momentum operator $\hat{p}_x=-i\hbar(\partial/\partial x)$ known from
quantum mechanics (for the momentum, see also section \ref{Fisher}).

In the usual approach, the momentum operator $\hat{p}_x=-i\hbar(\partial/\partial x)$
and the corresponding Hilbert space of the wave functions are postulated.
Then, the commutation relation $[x,\hat{p}_x]=i\hbar$ appears to be a
rather trivial mathematical identity. 
However, our approach is different.
We do not postulate the form of the momentum operator here and, instead it, we show that
the commutation relation (\ref{apx3}) containing the operator 
$-i(\partial/\partial x)$ appears in the probabilistic
description as a simple consequence
of the integration by parts applied to the normalization condition
(\ref{cnorm}) and the Born rule (\ref{Born}).
    
We note also that the second integral in Eq. (\ref{apx2}) does not depend on $s_1$ 
and shows that this commutation relation is the relation 
for $s_2=-(1/2)\ln\rho$ only.  
Therefore, existence of the commutation relation (\ref{apx3}) is related to
the existence of the probability distribution $\rho$.
In classical mechanics, where the probability distribution $\rho$
disappears, this commutation relation disappears, too.
It does not influence the classical dynamics which
depends only on the function $S_1=\hbar s_1$ playing the role of  
the classical action $S$ (see section \ref{classical}). 

It is seen that the usual quantization based on the transition 
from the classical coordinates and momentum to the coordinate 
and momentum operator obeying 
the commutation relations $[x,\hat{p}_x]=i\hbar$
corresponds to assuming the probabilistic character 
of measurements described by $\rho$.
Thus, discussion in sections \ref{Bornrule}
and \ref{commutation} helps to understand 
the postulates of quantum mechanics 
and clarifies the basic ideas of quantum mechanics.  
  
We note that similar commutation relations can be expected to be valid in any
probability theory formulated analogously to that discussed above.
In the limit case when the probabilistic distribution of the results
of measurements can be neglected (as, for example, in classical
mechanics) such commutation relations disappear.

\section{Uncertainty relations}

\label{uncertainty}
  
The uncertainty relation for the coordinate $x$ and the
operator $-i(\partial/\partial x)$ can be derived in a standard way
from the commutation relation (\ref{apx3}) (see e.g. \cite{Davydov,Shankar}).
Instead the standard approach, we will use results obtained above and
calculate $(v,v)$ in Eq. (\ref{vv}) for $\rho=|\psi|^2$ 
\be
\label{vv1}
(v,v)=\int\frac{1}{|\psi|^2}\left(\frac{\partial\rho}{\partial
  x}\right)^2{\rm d}V
=4\int\frac{1}{|\psi|^2}
\left[{\rm Re}
\left(\psi^*\frac{\partial\psi}{\partial x}\right)\right]^2{\rm d}V
=4\int\left(\frac{\partial s_2}{\partial x}\right)^2
{\rm e}^{-2s_2}{\rm d}V.
\ee
Using this result in Eq. (\ref{Schwarz}) for $n=0$
we get the uncertainty relation in the form 
\be
\label{ur1}
\int x^2 |\psi|^2 {\rm d}V \int\frac{1}{|\psi|^2}
\left[{\rm Re}
\left(\psi^*\frac{\partial\psi}{\partial x}\right)\right]^2{\rm d}V
\ge\frac{1}{4}.
\ee
It follows from Eq. (\ref{vv1}) that the last integral in Eq. (\ref{ur1}) depends 
only on the function $s_2$ giving the form of the probability 
distribution $\rho$ and does not depend on $s_1$. 
Thus, in contrast to the usual uncertainty relations in quantum mechanics, 
uncertainty relation (\ref{ur1}) does not depend on $s_1$. 

Now we compare the last uncertainty relation with the usual uncertainty relations.
First we see that
\be
\label{ur2}
\int\left(\frac{\partial s_2}{\partial x}\right)^2
{\rm e}^{-2s_2}{\rm d}V
\le
\int\bigg[\left(\frac{\partial s_1}{\partial x}\right)^2
+\left(\frac{\partial s_2}{\partial x}\right)^2\bigg]
{\rm e}^{-2s_2}{\rm d}V
=\int\left|-i\frac{\partial\psi}{\partial x}\right|^2{\rm d}V.
\ee
Substituting the right-hand side of this inequality into  
Eq. (\ref{Schwarz}) for $n=0$ we get the uncertainty relation
which looks more familiar than Eq. (\ref{ur1})
\be
\label{cSchwarz1}
\int x^2 |\psi|^2 {\rm d}V\int
\left|-i\frac{\partial\psi}{\partial x}\right|^2{\rm d}V\ge \frac{1}{4}.
\ee
It is seen from Eq. (\ref{ur2}) 
that this more usual form of the uncertainty relation depends both on $s_1$ and $s_2$
and the left hand side of Eq. (\ref{cSchwarz1}) is larger 
than or equal to that of Eq. (\ref{ur1}).

These results can be further generalized and even more general forms of the uncertainty
relations can be obtained.
Using the integration by parts and the condition $x|\psi|^2\rightarrow 0$
for $x\rightarrow\pm\infty$, 
Eq. (\ref{basic}) can be generalized as
\be
\label{apx4}
\int [(x-a)\psi]^*
\bigg[\frac{\partial\psi}{\partial x}-i b\psi\bigg] {\rm d} V
+\int\bigg[\frac{\partial\psi}{\partial x}-
i b\psi\bigg]^*[(x-a)\psi] {\rm d} V=-1,
\ee
where $a$ and $b$ are arbitrary real constants.
This equation can be written as
\be
\label{auv}
(u,v)+(v,u)=-1,
\ee
where 
\be
\label{aux}
u=(x-a)\psi
\ee
and 
\be
\label{avx}
v=\frac{\partial\psi}{\partial x}-ib\psi.
\ee
Using the property $(v,u)=(u,v)^*$ we get 
\be
2\,{\rm Re}(u,v)=-1.
\ee
Calculating the square of the last equation we get successively
\be
\label{aineq}
1=4 [{\rm Re}(u,v)]^2 \leq 4\{[{\rm Re}(u,v)]^2+[{\rm Im}(u,v)]^2\}=4|(u,v)|^2.
\ee
Using this result and the Schwarz inequality (\ref{Schwarzorig})
we obtain the inequality
\be
\label{auuvv}
(u,u)(v,v) \geq \frac{1}{4}
\ee
that can be rewritten in form of the uncertainty relation  
\be
\label{auuvv2}
\int (x-a)^2 |\psi|^2\, {\rm d}V \int \bigg|
-i\frac{\partial\psi}{\partial x}-b\psi\bigg|^2 {\rm d}V \geq \frac{1}{4}.
\ee
This general form of the uncertainty relation
is valid for any real numbers $a$ and $b$.

Interesting question is to find the values of $a$ and $b$ leading
to the smallest value of the left hand side of the last uncertainty relation. 
The minimum of the left hand side of Eq. (\ref{auuvv2}) is obtained for
\be
\label{a}
a=\int \psi^* x \psi\, {\rm d} V=\langle x\rangle
\ee
and
\be
\label{b}
b=\int \psi^*\left(-i\frac{\partial\psi}{\partial x}\right){\rm d} V=
\left\langle -i\frac{\partial}{\partial x} \right\rangle.
\ee
Except for the factor $\hbar^2$, the resulting uncertainty relation
\be
\label{cSchwarz2}
\int (x-\langle x\rangle)^2 |\psi|^2 {\rm d}V
\int\bigg|-i\frac{\partial\psi}{\partial x}
-\left\langle -i\frac{\partial}{\partial x}\right\rangle\psi\bigg|^2{\rm d}V
\ge \frac{1}{4}
\ee
agrees with the well-known Heisenberg uncertainty relation \cite{Heisenberg,Davydov,Shankar}
\be
\label{Heiss}
\int (x-\langle x\rangle)^2 |\psi|^2 {\rm d}V
\int\bigg|-i\hbar\frac{\partial\psi}{\partial x}
-\left\langle -i\hbar\frac{\partial}{\partial x}\right\rangle\psi\bigg|^2{\rm d}V
\ge \frac{\hbar^2}{4}.
\ee
Therefore, the Heisenberg uncertainty relation corresponds to the smallest
value of the left hand side of a more general uncertainty relation (\ref{auuvv2})
in which both the functions $s_1$ and $s_2$ are taken into account.  

Now we want to clarify the question whether it is possible to make the left hand side of
the uncertainty relation (\ref{cSchwarz2}) even smaller.
First we see that the second integral in the last uncertainty relation is greater
than or equal to the integral appearing in Eq. (\ref{vv1})
\be
\int\bigg|-i\frac{\partial\psi}{\partial x}
-\left\langle -i\frac{\partial}{\partial x}\right\rangle\psi\bigg|^2{\rm d}V
=
\ee
\bd
=\int\bigg[\left(\frac{\partial s_1}{\partial x}\right)^2
+\left(\frac{\partial s_2}{\partial x}\right)^2\bigg]
{\rm e}^{-2s_2}{\rm d}V
-\bigg(\int\frac{\partial s_1}{\partial x}
{\rm e}^{-2s_2}{\rm d}V\bigg)^2
\ge\int\left(\frac{\partial s_2}{\partial x}\right)^2
{\rm e}^{-2s_2}{\rm d}V.
\ed
Here, the equality is obtained only if the function $s_1$ does not
depend on $x$.
In deriving the preceding relation we used the result
\be
\int\frac{\partial s_2}{\partial x}{\rm e}^{-2s_2}{\rm d}V=
-\frac{1}{2}\int\frac{\partial {\rm e}^{-2s_2}}{\partial x}{\rm d}V=
-\frac{1}{2}\int\frac{\partial|\psi|^2}{\partial x}{\rm d}V=
-\frac{1}{2}\int|\psi|^2\bigg|_{x=-\infty}^{\infty}{\rm d}y{\rm d}z=0.
\ee

Further, we can repeat the procedure used in section
\ref{Bornrule} with $u=(x-\langle x\rangle)\,\psi$ and obtain
the uncertainty relation
\be
\int (x-\langle x\rangle)^2 |\psi|^2 {\rm d}V \int\frac{1}{|\psi|^2}
\left[{\rm Re}
\left(\psi^*\frac{\partial\psi}{\partial x}\right)\right]^2{\rm d}V
\ge\frac{1}{4}
\ee
which is generalization of the uncertainty relation (\ref{ur1}).
The left hand side of this uncertainty relation is less than or equal to
that of Eq. (\ref{cSchwarz2}) and the equality is obtained for $s_1$ independent
of $x$. 
Therefore, if the function $s_1$ depends on $x$, the Heisenberg uncertainty relation (\ref{Heiss})
can be replaced by the uncertainty relation having smaller value of its left hand side
\be
\label{Heiss2}
\int (x-\langle x\rangle)^2 |\psi|^2 {\rm d}V
\int\frac{1}{|\psi|^2}
\left\{{\rm Re}
\left[\psi^*\left(-i\hbar\frac{\partial\psi}{\partial x}\right)\right]\right\}^2{\rm d}V
\ge \frac{\hbar^2}{4}.
\ee

We note that, similarly to the commutation relations, these uncertainty relations are
straightforward consequence of the integration by parts applied 
to the definition of the mean value (\ref{meanx}) 
and the Born rule (\ref{Born}).
It is also seen that similar uncertainty relations can be obtained not
only in quantum mechanics but in any probabilistic
theory of a similar form.
If the probabilistic character of measurements can be neglected
(as for example in case of classical mechanics),
the uncertainty relations disappear analogously to the commutation relations. 

Another discussion of the uncertainty relations can be found 
for example in \cite{Braunstein,Brody,Hall2}.

There are two important operators appearing in the
commutation and uncertainty relations discussed above:  
the coordinate $x$ and the operator $-i(\partial/\partial x)$.
Except for $\hbar$, the operator $-i(\partial/\partial x)$ 
equals the momentum operator $\hat{p}_x=-i\hbar(\partial/\partial x)$ known
from quantum mechanics.
The momentum operator will be discussed in more detail in section
\ref{Fisher}.

\section{Vector potential}
\label{vector}

It is worth noting that Eq. (\ref{apx4}) remains valid also in case 
that the constant $b$ is replaced by a real function $b=f_x({\bf r},t)$.
It means that the operator $-i(\partial\psi/\partial x)$ can
be replaced by the operator $-i(\partial\psi/\partial x)-f_x$
and the commutation relation (\ref{apx3}) and
the uncertainty relation (\ref{auuvv2}) can be further generalized.
Therefore, general structure of the
theory remains preserved for any real function $f_x$.
In physics, the functions $f_x$, $f_y$ and $f_z$ can 
for example correspond to the
components of the electromagnetic vector potential ${\bf A}=(A_x,A_y,A_z)$
multiplied by the charge $q$ of the particle.
Except for $\hbar$, it agrees with the rule 
$-i\hbar\nabla\rightarrow -i\hbar\nabla-q{\bf A}$
for including the vector potential ${\bf A}$ into quantum theory
(for charge, see the end of section \ref{time}).

It is seen that possibility to include the vector potential into the theory
is another general consequence of the definition (\ref{meanx}), 
integration by parts and the Born rule (\ref{Born}).

\section{Probability density current}

\label{probability}

To describe physical systems,
we have to specify not only the form of the probability distribution 
$\rho=\rho({\bf r},t)$ determining the mean values 
$\langle x^n\rangle$ at time $t$ but also the motion of the measured system in space.
Information about the motion of the system in space can be described 
by means of the function $s_1$ in Eq. (\ref{psi})
which could be an arbitrary real function till now.
Here, we can proceed similarly as in continuum mechanics,
where not only the density $d$ of the continuum but also
the corresponding density current ${\bf j}=d\,{\bf v}$, where ${\bf v}$ 
is the velocity of the continuum is defined.

Since we use the probability density $\rho$ here, 
the corresponding current will have meaning of the 
probability density current.
Analogously to the continuum mechanics, we can define 
the probability density current $j_k$ by the equation  
\be
\label{j_k}
j_k=\rho\, v_k, \quad k=1,2,3
\ee
where the vector $v_k$ does not give the real velocity of the particle 
and has only the probabilistic meaning.
Further, we write the vector $v_k$ by means of the gradient of the function $s_1$ 
\be
\label{v_k}
v_k= \frac{\hbar}{m_0}\frac{\partial s_1}{\partial x_k},
\ee
where $m_0$ is the mass of the particle and 
$\hbar$ is a constant that determines the units used in measurement 
(see also \cite{Reginatto}).
The numerical value of $\hbar$ has to be found experimentally.

Equation (\ref{v_k}) can be viewed as a probabilistic generalization of
the classical expression ${\bf v}={\bf p}/m_0=(\nabla S)/m_0$, where ${\bf p}$ is
the classical momentum and $S$ is the classical action.
One can be also inspired by the eikonal theory or 
take Eq. (\ref{v_k}) as a purely mathematical trick 
leading to a simple mathematical expression between $j_k$ and $s_1$.
The spatial derivative in Eq. (\ref{v_k}) could be also replaced by a different 
functional relation between $j_k$ and $s_1$.
However, use of the spatial derivative in Eq. (\ref{v_k}) has two important advantages:
It leads to the linearity of the expression for $j_k$ in terms of $\psi$ discussed below 
and contains the spatial derivatives $(\partial/\partial x_k)$ 
that appeared in the commutation and uncertainty relations
discussed in the preceding sections.
Thus, formulation based on Eqs. (\ref{j_k}) and (\ref{v_k}) 
has important mathematical and physical advantages and contains the information on 
the motion of the measured systems in space. 
We note also that the number of the quantities $\rho$ and $j_k$, $k=1,2,3$ 
equals the number of the space-time dimensions.    

Using Eqs. (\ref{j_k})-(\ref{v_k}) we get successively
\be
\label{j1}
j_k=\rho\,v_k=\rho\frac{\hbar}{m_0}\frac{\partial s_1}{\partial x_k}=
\sqrt{\rho}{\rm e}^{-is_1}\,\sqrt{\rho}{\rm e}^{is_1}
\frac{\hbar}{m_0}\frac{\partial s_1}{\partial x_k}
=\frac{\hbar}{m_0}\left[\sqrt{\rho}{\rm e}^{-is_1}(-i)\frac{\partial(\sqrt{\rho}{\rm e}^{is_1})}{\partial x_k}
+\frac{i}{2}\frac{\partial \rho}{\partial x_k}\right].
\ee
Now, we can use the complex probability amplitude (\ref{psi})
and get
\be
\label{j2}
j_k=\frac{\hbar}{m_0}\left[\psi^*\left(-i\frac{\partial\psi}{\partial x_k}\right)
+\frac{i}{2}\frac{\partial \rho}{\partial x_k}\right].
\ee
However, the probability density current is real.
Therefore, by calculating the real part of the last expression we obtain 
the final formula for $j_k$
\be
\label{j3}
j_k=
\frac{\hbar}{2m_0}\left[\psi^*\left(-i\frac{\partial \psi}{\partial
    x_k}\right)+c.c.\right]
=\frac{\hbar}{2m_0i}\left(\psi^*\frac{\partial \psi}{\partial
    x_k}-\psi\frac{\partial \psi^*}{\partial x_k}\right).
\ee
This formula agrees
with the expression for the probability density current
known from quantum mechanics \cite{Davydov,Shankar}. 

The manner of deriving Eq. (\ref{j3}) shows that
the relation between the probability amplitude and
probability density in form of the Born rule (\ref{Born})
yields not only physically meaningful quantities in 
the uncertainty relations (\ref{Schwarz}) 
but leads also to a simple expression for
the probability density current (\ref{j3}).  
 
It is seen from Eq. (\ref{j3}) that to obtain nonzero $j_k$,  
the probability amplitude $\psi$ must be complex.
In agreement with the rules known from quantum mechanics, 
the probability amplitudes $\psi$ and $\psi\exp(i\alpha)$, where $\alpha$
is a real constant, yield the same probability density $\rho$
and probability density current $j_k$.

It is to be noted that the expression for the probability amplitude in the form
\be
\label{psi12}
\psi={\rm e}^{is}={\rm e}^{is_1-s_2}
\ee
describes two different aspect of measurements:
The first one represented by the function $s_2$ is related to the probability density 
$\rho$ via the equation $s_2=-(1/2)\ln\rho$.
The second one represented by $s_1$ is related to the probability density
current ${\bf j}=\rho\hbar\nabla s_1/m_0$.
In other words, they are two different kinds of physical information carried  
by two different functions $s_1$ and $s_2$.
In this sense, the probability amplitude $\psi=\exp(is_1-s_2)$ represents 
the state of the system.

\section{Fisher information, kinetic energy and momentum operator}
\label{Fisher}

We note that the integral $(v,v)$ in Eq. (\ref{vv}) which appears in
the first uncertainty relation discussed in section 
\ref{uncertainty} (see Eqs. (\ref{vv1})-(\ref{ur1}))
\be
I_x=(v,v)=\int\frac{1}{\rho}\left(\frac{\partial\rho}{\partial x}\right)^2
{\rm d}V=4\int\left(\frac{\partial|\psi|}{\partial x}\right)^2{\rm d}V
=4\int\rho\left(\frac{\partial s_2}{\partial x}\right)^2{\rm d}V
\ee 
is known as the Fisher information characterizing
the probability distribution $\rho$ with respect to the variable $x$
\cite{Fisher,Frieden,Cover,Hall1,Parwani}.
The following inequalities can be obtained from Eq. (\ref{Schwarz})
\be
\label{boundx}
\langle x^2\rangle\ge \frac{1}{I_x}
\ee
and
\be
\label{boundx2}
\langle x^{2n+2}\rangle\ge \frac{[(n+1)\langle x^n\rangle]^2}
{I_x}, \quad n=1,2,3,\ldots.
\ee
Therefore, the Fisher information $I_x$ has simple physical meaning: 
It determines the lower bounds to the quantities $\langle x^{2n+2}\rangle$,
$n=0,1,2,\ldots$ characterizing results of measurements.
The larger value of $I_x$, the smaller values of
$\langle x^{2n+2}\rangle$, $n=0,1,2,\ldots$ can be obtained in measurement.
In three dimensions, the corresponding Fisher information can be written as
\be
I=I_x+I_y+I_z=4\int\rho(\nabla s_2)^2{\rm d}V.
\ee 

However, these Fisher informations 
depend on $s_2$ only and do not 
take into consideration the probability current 
represented by $s_1$.
Therefore, the Fisher informations $I_x$ and $I$
can be in physics generalized as \cite{Frieden}
\be
I'_x
=4\int\rho\bigg[\left(\frac{\partial s_1}{\partial x}\right)^2
+\left(\frac{\partial s_2}{\partial x}\right)^2\bigg]{\rm d}V
=4\int\left|-i\frac{\partial \psi}{\partial x}\right|^2{\rm d}V
\ge I_x
\ee 
and
\be
\label{IQM}
I'=I'_x+I'_y+I'_z=4\int\rho\left[\nabla s_1)^2+(\nabla s_2)^2\right]{\rm d}V
=4\int|-i\nabla\psi|^2 {\rm d}V\ge I.
\ee

The last integral appears also in
the expression for the kinetic energy known from quantum mechanics
\be
\label{T}
T=\frac{\hbar^2}{2m_0}\int|-i\nabla\psi|^2 {\rm d}V=
\frac{\hbar^2I'}{8m_0}. 
\ee
Therefore, the kinetic energy $T$ in quantum
mechanics is proportional to
the generalized Fisher information $I'$ in which
both the form of the probability density $\rho$ and the motion
of the measured system in space are taken into account.

In classical mechanics, the probability density
$\rho({\bf r},t)$ is as narrow that it can be replaced by the function 
$\delta({\bf r}-\langle {\bf r}\rangle)$, 
where $\langle {\bf r}\rangle=\langle {\bf r}\rangle(t)$
is the classical trajectory.  
Then, integration in Eq. (\ref{T})
disappears and the kinetic energy depends only on 
the function $S_1(\langle {\bf r}\rangle,t)=\hbar s_1(\langle {\bf r}\rangle,t)$ 
playing the role of the classical action $S$
(see section \ref{classical}). 

The uncertainty relation (\ref{auuvv2}) can be 
for $a=\langle x\rangle$ and $b=0$ written in the form 
\be
\label{boundx'}
\langle(x-\langle x\rangle)^2\rangle\ge \frac{1}{I'_x}
\ee
which is more general than Eq. (\ref{boundx}).
In classical mechanics, the kinetic energy related to $s_1$ 
and the Fisher information
$I'$ are very large and the mean square displacement
$\langle (x-\langle x\rangle)^2\rangle$ is as small that
the classical trajectories can be introduced.   

At the end of this section, we will make a few remarks on
the momentum operator.
In contrast to standard quantum mechanics, 
the hermitian operator $-i\hbar\nabla$ need not be postulated
in our approach.
Its appearance in the commutation
relations, uncertainty relations,
probability density current and
the generalized Fisher information 
indicates its important role in the theory. 
Further, relation (\ref{j3}) between the probability density current ${\bf j}$ 
and the operator $-i\hbar\nabla$ shows 
that this operator can be used for describing the motion of the measured system in space.
It agrees with quantum mechanics, where 
the operator $\hat{\bf p}=-i\hbar\nabla$
is postulated as the momentum operator. 
Concrete physical meaning of the operator $\hat{\bf p}$ --- 
the momentum operator --- can be clarified in different ways.
Probably the best approach is based on the transition to
classical mechanics discussed in section \ref{classical}.
Performing this transition, the quantity
$\int|-i\hbar\nabla \psi|^2{\rm d}V/(2m_0)$ 
appearing in the Schr\"odinger equation becomes the kinetic energy 
in the Hamilton-Jacobi equation 
$(\nabla S)^2/(2m_0)$,
where $S$ is the classical action (see section \ref{classical}).
Taking into account that  ${\bf p}=\nabla S$ is the classical
momentum, the operator $\hat{\bf p}=-i\hbar\nabla$
can be denoted as the momentum operator.

\section{Probability density and time}

\label{time}

Time can be discussed analogously to the spatial coordinates, however,
there are some important differences that have to be respected.

Assuming that the initial conditions for $\psi$ at time $t=0$ are given,
the probability amplitude $\psi({\bf r},t)$, $t>0$ gives the probabilistic 
description of measurements made at later times.
Therefore, time evolution has unidirectional character from given
initial conditions to the relative probability of results of (yet unperformed)
measurements at later times.
If such measurement is actually performed, this probabilistic description must
be replaced by a concrete result obtained from the performed
measurement.
This is the basis of two different parts of the evolution scheme in quantum
mechanics: 
(1) Evolution between the initial conditions at $t=0$ and time $t>0$ 
of the following measurement. 
This time evolution is described by the equations of motion. 
(2) Reduction or collapse of the wave function at time $t$ of the actually performed 
measurement.
In this paper, we are interested mainly in equations of motion,
i.e. in the first part of this evolution scheme.
Detailed microscopic description of the reduction of the probability amplitude
is not needed here and can be found for instance 
in \cite{Allah1,Allah2}, see also \cite{Klyshko}.

In case of the spatial coordinates discussed in section \ref{Bornrule}, 
we investigated the bound states obeying the normalization condition 
\be
\label{norm1}
\int|\psi({\bf r},t)|^2{\rm d}V=1
\ee
valid at all times,
i.e. we assumed the infinite life time of the investigated system.
We assumed also that the mean values
\be
\langle x^n\rangle=\int x^n|\psi({\bf r},t)|^2{\rm d}V
\ee 
are finite.

In case of time, we would like to proceed similarly as for 
the spatial coordinates.
We would like to introduce the mean values
\be
\langle t^n\rangle=\int_0^{\infty}\int t^n|\psi({\bf r},t)|^2{\rm d}V{\rm d}t,
\ee 
where, in agreement with our understanding of the arrow of time,
the time integration is carried out for $t\ge 0$.
Then, we want to derive the corresponding commutation and uncertainty relations
for time and energy, discuss the scalar potential, antiparticles and equations of motion
(sections \ref{time}--\ref{equations}).
 
However, taking the usual normalization condition (\ref{norm1}), i.e. 
assuming the infinite life time, 
the mean values $\langle t^n\rangle$ tend to infinity and similar
approach as in case of the spatial coordinates cannot be used.

The most simple solution of this problem is to assume that the life time
of the investigated system is finite and replace the probability amplitude $\psi$
by the function
\be
\label{chi}
\chi({\bf r},t)=\psi({\bf r},t)\eta(t).
\ee
Here, $\eta(t)$ is a real decaying function normalized
by the condition
\be
\int_0^{\infty}|\eta(t)|^2{\rm d}t=1
\ee
for which the integrals
\be
\label{meant}
\langle t^n\rangle=\int_0^{\infty}t^n|\eta(t)|^2{\rm d}t
\ee
have finite values
and $\psi({\bf r},t)$ obeys the condition (\ref{norm1}).
As a concrete example of a function obeying these
conditions we can take the exponential
\be
\label{eta}
\eta(t)=\frac{{\rm e}^{-t/(2\tau)}}{\sqrt{\tau}},\quad\tau>0
\ee
used often in quantum mechanics for describing the finite life time.

Then, repeating essentially the same procedure
as for the spatial coordinates, we can define
the square of the mean life time $\langle t^2\rangle$, 
get the operator $i(\partial/\partial t)$, 
obtain the corresponding time commutation and uncertainty relations 
and introduce the scalar potential.
After performing all calculations in the space-time region,
transition to standard quantum mechanics of
particles having the infinite life time can be done
by assuming that $\eta(t)$ changes very slowly in time or 
$\tau\rightarrow\infty$.
At the same time, the normalization condition 
over the space-time
\be
\label{norm}
\int_0^{\infty}\int|\chi({\bf r},t)|^2{\rm d}V{\rm d}t=1
\ee
has to be replaced
by the standard normalization condition (\ref{norm1}).

Because of the analogy of this approach 
to that for the spatial coordinates, we will present
here only the most important steps of this discussion.
By analogy with Eq. (\ref{psi12}), we write $\chi$
in the form
\be
\chi={\rm e}^{is_1-s_2}.
\ee  
Then, we define the time component of the probability density
current by the equation analogous to Eqs. (\ref{j_k})-(\ref{v_k})
\be
\label{j_t}
j_t=-\rho\frac{\hbar}{m_0}\frac{\partial s_1}{\partial t}
\ee
and obtain expression similar to Eq. (\ref{j3})
\be
j_t=\frac{\hbar}{2m_0}\left[\chi^*\left(i\frac{\partial \chi}{\partial t}\right)+c.c.\right].
\ee
We note that the sign in the definition of $j_t$ is opposite 
than in case of $j_k$.
It corresponds to different signs of the time and spatial parts of the metric
in the special relativity. 
Equation (\ref{j_t}) corresponds to the time component of the 
probability density current 
$j_0={\rm Re}[\psi^*i\hbar(\partial\psi/\partial x_0)]/m_0$ known
from relativistic quantum mechanics \cite{Davydov,Shankar},
where $x_0=ct$.

Further, by analogy with Eq. (\ref{apx2}), we get the equation
\be
\label{apt2}
\int_{t=0}^{\infty}\int\left[\left(i\frac{\partial\chi}{\partial t}\right)^*t\chi
-(t\chi)^*\left(i\frac{\partial\chi}{\partial t}\right) \right]{\rm d}V{\rm d}t
=i.
\ee
Similarly to the constant $b$ in Eq. (\ref{apx4}), 
we can introduce also a real constant $b'$ into this equation
\be
\label{apt3}
\int_{t=0}^{\infty}\int
\left[\left(i\frac{\partial\chi}{\partial t}-b'\chi\right)^*t\chi-\right.
\left.(t\chi)^*\left(i\frac{\partial\chi}{\partial t}-b'\chi\right) \right]{\rm d}V{\rm d}t
=i.
\ee
In contrast to the spatial coordinate $x$, where we could introduce an arbitrary 
shift $a$, see Eq. (\ref{apx4}), the time integration in the 
last equations runs from time $t=0$ when the initial conditions were given and
cannot be arbitrarily shifted.
The corresponding time uncertainty relation can be written in form analogous
to Eq. (\ref{auuvv2})
\be
\label{tuuvv2}
\int_0^{\infty}\int t^2 |\chi|^2\,{\rm d}V{\rm d}t 
\int_0^{\infty}\int\left|i\frac{\partial\chi}{\partial t}-b'\chi\right|^2 
{\rm d}V{\rm d}t \geq \frac{1}{4}.
\ee
The minimum of the left-hand side of Eq. (\ref{tuuvv2}) is obtained for
\be
\label{bmin}
b'=
\frac{1}{2}\left[\int_0^{\infty}\int \chi^*i\frac{\partial\chi}{\partial t} 
{\rm d}V{\rm d}t+c.c.\right].
\ee
We note that Eq. (\ref{tuuvv2}) is valid also if $b'$
is replaced by a real function $f_0({\bf r},t)$.

For $b'=0$ we get the uncertainty relation 
\be
\label{t^2I}
\langle t^2\rangle\ge \frac{1}{I''_t}
\ee
between the mean square time 
\be
\langle t^2\rangle=\int_0^{\infty}\int t^2 |\chi|^2\,{\rm d}V{\rm d}t 
\ee
and the time Fisher information
\be
I''_t=4\int_0^{\infty}\int\left|i\frac{\partial\chi}{\partial t}\right|^2 
{\rm d}V{\rm d}t,
\ee
where the symbol $''$ denotes integration over the space-time.
It is seen from Eq. (\ref{t^2I}) that the time Fisher information $I''_t$ gives the lower bound
to the mean square time.

To illustrate the meaning of the uncertainty relation (\ref{tuuvv2}) 
we consider the decaying probability amplitude
\be
\chi({\bf r},t)=
\frac{1}{\sqrt{\tau}}{\rm e}^{-i\omega t-t/(2\tau)}\psi({\bf r}),
\ee
where the spatial part of the probability amplitude $\psi$ is normalized
by the usual condition $\int|\psi({\bf r})|^2{\rm d}V=1$.
In this case, we get from Eqs. (\ref{tuuvv2})-(\ref{bmin})
\be
\int_0^{\infty}\int t^2 |\chi|^2\,{\rm d}V{\rm d}t=2\tau^2,
\ee
$b'=\omega$ and
\be
\int_0^{\infty}\int\left|i\frac{\partial\chi}{\partial t}-b'\chi\right|^2 
{\rm d}V{\rm d}t=\frac{1}{4\tau^2}.
\ee
Therefore, inequality (\ref{tuuvv2}) gives
the relation between the mean square time 
$\langle t^2\rangle=2\tau^2$
and the square of the absolute value of the imaginary part 
of the complex frequency $\omega-i/(2\tau)$
and has the physical meaning of the well-known time--energy uncertainty 
relation (see e.g. \cite{Davydov,Shankar,Busch}).

Now, we will discuss different roles of time and spatial coordinates 
in quantum mechanics (see also \cite{Hilgerwood}). 
In quantum mechanics, it is assumed that the system is prepared 
in a state given by the initial conditions for the
wave function $\psi({\bf r},t)$ in the whole space at time $t=0$.
Then, equations of motion are used to calculate the wave function
at later times and to determine the relative probability of results of
future measurements made on the system.
For this reason, the spatial coordinates and time have different
roles in quantum mechanics and it is not surprising that 
the coordinate--momentum and time--energy uncertainty relations have
different character.
The former ones give the relation between the mean square deviations 
of the coordinate and momentum from their mean values and can be used
for the infinite as well as the finite life time.
The latter uncertainty relation is the relation between the mean life time 
(no mean square deviation) and the imaginary part of the energy
(i.e. the width of the corresponding energy level) and can be 
used for system having the finite life time only.

\section{Scalar potential and antiparticles}
\label{scalar}

In agreement with our understanding of the arrow of time
from $t=0$ to $t>0$, 
we assume that direct physical meaning have only 
the probability amplitudes corresponding to the non-negative values
of the time component of the probability density current integrated over
the whole space
\be
\label{j_t0}
\int j_t{\rm d}V=-\frac{\hbar}{m_0}\int\rho\frac{\partial s_1}{\partial t}{\rm d}V\ge 0.
\ee
If this quantity is negative, its sign can be reversed by 
the complex conjugation $\chi\rightarrow\chi^*$ changing the sign
of the phase $s_1$ and the probability density currents $j_k$ and $j_t$. 
Performing this transformation we get from Eq. (\ref{apt3})
for $b'=f_0({\bf r},t)$
\be
\label{apt7}
\int_{t=0}^{\infty}\int
\bigg[\bigg(i\frac{\partial\chi}{\partial t}+f_0\chi\bigg)^*t\chi-
(t\psi)^*\bigg(i\frac{\partial\chi}{\partial t}+f_0\chi\bigg) \bigg]
{\rm d}V{\rm d}t=i
\ee
and see that it leads to the change of the sign of the function $f_0$.

Analogous discussion can be done also for the spatial coordinates.
As a result, the complex conjugation $\chi\rightarrow\chi^*$ 
or $\psi\rightarrow\psi^*$ leads
to the change of sign of the functions $f_0$ and $f_k$, $k=1,2,3$,
that can be respected by putting $f_0=qU$ and $f_k=qA_k$, where
$U$ and $A_k$ can be for example the scalar and vector electromagnetic potentials
and $q$ denotes the charge of the particle.
Therefore, the probability amplitudes $\psi$ 
and $\psi^*$ describe particles that differ by the sign of their 
charge and general structure of our probabilistic description and the unidirectional
character of time contribute to understanding the 
existence of particles and antiparticles.

Except for $\hbar$, these conclusions agree with the well-known rules 
$i\hbar(\partial/\partial t)\rightarrow i\hbar(\partial/\partial t)-qU$
and $-i\hbar\nabla\rightarrow -i\hbar\nabla-q{\bf A}$
for including the electromagnetic potentials into quantum theory.
These potentials representing different physical scenarios 
do not appear among the variables of the probability amplitude
and describe non-quantized classical fields.

We have seen  that to obtain results of sections
\ref{Bornrule}-\ref{scalar}
no equations of motion have been needed.
Thus, this part of the mathematical formalism of quantum mechanics
follows directly from the probabilistic description of results of measurements.
The Planck constant determines the units used in
measurements and scales at which
the probabilistic character of measurements is important.
It must be determined experimentally.

\section{Equations of motion}

\label{equations}

First we note that the physical content of the Fisher 
information --- characterization of the smoothness of the probability distribution ---
is similar to that of the Shannon entropy.
However, in contrast to the Shanon entropy, the Fisher information
depends on the local properties of the probability distribution
and can be used for deriving equations of motion \cite{Frieden}.

To find equations of motion we will require
relativistic invariance of the theory.
In this respect, our approach is different from that based
on the principle of extreme physical information \cite{Frieden} 
or minimum of the Fisher information \cite{Reginatto}.
Except for the end of this section, 
we will discuss free particles with no external fields.  

To find relativistically invariant formulation, 
we use four time and spatial Fisher informations
\be
I''_t=4\int_0^{\infty}\int\left|i\frac{\partial\chi}{\partial t}\right|^2 
{\rm d}V{\rm d}t,
\ee 
\be 
I''_x=4\int_0^{\infty}\int\left|-i\frac{\partial\chi}{\partial x}\right|^2 
{\rm d}V{\rm d}t,\quad
I''_y=4\int_0^{\infty}\int\left|-i\frac{\partial\chi}{\partial y}\right|^2 
{\rm d}V{\rm d}t,\quad
I''_z=4\int_0^{\infty}\int\left|-i\frac{\partial\chi}{\partial z}\right|^2 
{\rm d}V{\rm d}t
\ee
giving the lower bounds to $\langle t^2\rangle$, $\langle(x-\langle x\rangle)^2\rangle$, 
$\langle(y-\langle y\rangle)^2\rangle$ and $\langle(z-\langle z\rangle)^2\rangle$
(see Eqs. (\ref{boundx'}) and (\ref{t^2I})). 
In contrast to the integration over all
times used in \cite{Frieden, Reginatto}, 
the time integration is performed here over the physically relevant region from 
the initial conditions at $t=0$ to infinity. 

Due to the possibility of taking arbitrary physically reasonable 
initial conditions for $\chi({\bf r},t)$ at $t=0$, 
the spatial Fisher informations can have arbitrary
values in the region
\be
\label{I''pos}
I''_x\ge 0,\quad I''_y\ge 0,\quad I''_z\ge 0.
\ee
In this sense, the spatial Fisher informations are independent quantities.
The time Fisher information is also non-negative, $I''_t\ge 0$, however,
its value is given not only by the initial conditions for $\chi({\bf r},t)$ 
at $t=0$ but also by the requirement of the relativistic invariance of 
the combined Fisher informations discussed below.
From this point of view, the time Fisher information $I''_t$ is not
an independent quantity.
      
To create the relativistic invariant from the Fisher informations,
we will first take their linear combination in general form
\be
\label{I''a}
\frac{I''_t}{c^2}\pm(I''_x+I''_y+I''_z)=const
\ee
and assume that it has a value independent of the inertial system
in which the measurement is performed. 
Here, $c$ is the speed of light 
and we assume first that the plus or minus sign can be taken in this equation. 

We note that for a very flat distribution $\rho$ and 
the probability density current $j_k=0$ the spatial Fisher informations
approach zero and the time Fisher information equals $I''_t=c^2\,const$ 
(we say usually that the free particle is in its rest frame).
It follows from here and the condition $I''_t\ge 0$ that the constant
$const$ must be greater than or equal to zero
\be
const\ge 0.
\ee

Further we note that going from the inertial coordinate system in which $j_k=0$ 
to another one where $j_k\neq 0$ (the particle moves with respect to this system)
the sum of the spatial Fisher informations $I''_x+I''_y+I''_z$ 
(and the kinetic energy of the particle) increases.
Assuming first the positive sign in front of the sum of the spatial Fisher informations,
we see that for $I''_x+I''_y+I''_z >const$ the time Fisher information $I''_t$
would become negative in contradiction with its property $I''_t\ge 0$. 
Therefore, to guarantee that the left hand side of Eq. (\ref{I''a}) equals $const\ge 0$
and $I''_t\ge 0$ for all values of $I''_x+I''_y+I''_z\ge 0$, 
we have to take the time and spatial Fisher informations
with a different sign and assume
\be
\label{I''}
\frac{I''_t}{c^2}-(I''_x+I''_y+I''_z)=const.
\ee
Thus, our analysis based on the properties of the Fisher informations
confirms that the signs of the metric used in Eq. (\ref{I''}) and the special theory of relativity
must be different for time and the spatial coordinates.
We note that, in contrast to the constant $const$,
the Fisher informations $I''_t$, $I''_x$, $I''_y$ and $I''_z$ can have different values
in different inertial systems. 

Rewriting Eq. (\ref{I''}) into the form
\be
\label{E^2a}
\int_0^{\infty}\int\left|i\hbar\frac{\partial\chi}{\partial t}\right|^2 
{\rm d}V{\rm d}t=
c^2\int_0^{\infty}\int\left|-i\hbar\nabla\chi\right|^2 
{\rm d}V{\rm d}t+\frac{const\,\hbar^2c^2}{4}
\ee
it can be compared with the well-known Einstein equation
\be
\label{EE}
E^2=c^2p^2+m_0^2c^4.
\ee
We will consider a free particle described by the probability
amplitude 
\be
\label{free}
\chi({\bf r},t)={\rm e}^{(Et-{\bf p}{\bf r})/(i\hbar)}
\frac{{\rm e}^{-t/(2\tau)}}{\sqrt{\tau}}
\left(\frac{\alpha}{\pi}\right)^{3/4}{\rm e}^{-\alpha r^2/2},\quad \tau>0,\,\alpha>0,
\ee
where $\hbar$ depends on the units used in measurement.
Here, the first exponential represents the probability amplitude describing
a free particle with the same probability density to find the particle anywhere
in time and space, $E$ and ${\bf p}$ are a real number and a real vector.
To obey the bound state normalization condition (\ref{norm}), we multiplied the first exponential
by two additional ones corresponding to a very long life time $\tau$ 
and a very small spatial damping factor $\alpha$.
Calculating all integrals in Eq. (\ref{E^2a}) and putting $\tau\rightarrow\infty$ and
$\alpha\rightarrow 0$ at the end of the calculation we get 
\be
\label{E^2}
E^2=c^2p^2+\frac{const\,\hbar^2\,c^2}{4}.
\ee
It is seen that this equation agrees with Eq. (\ref{EE}) for 
\be
\label{const}
const=\frac{4m_0^2c^2}{\hbar^2}.
\ee   
This discussion shows again
that the operators $i\hbar(\partial/\partial t)$
and $-i\hbar\nabla$ in Eq. (\ref{E^2a}) can be denoted as the energy and momentum
operators, respectively.
   
In the rest of this section, we will derive the most important 
equations of motion of quantum mechanics in a similar way as 
in \cite{Frieden}. 
Using the normalization condition (\ref{norm}) we 
rewrite Eq. (\ref{I''}) into the form
\be
\label{Kvv1}
J[\chi]=\int_0^{\infty}\int\bigg(
\frac{1}{c^2}\bigg|\frac{\partial\chi}{\partial t}\bigg|^2
-|\nabla\chi|^2-\frac{const}{4}|\chi|^2\bigg){\rm d}V{\rm d}t=0.
\ee
Equation of motion can be found from the condition 
that $J[\chi]$ does not depend on $\chi$
\be
\delta J[\chi]=0
\ee
or
\be
\label{Kvv2}
\int_0^{\infty}\int\bigg(
\frac{1}{c^2}\frac{\partial\delta\chi^*}{\partial t}\frac{\partial\chi}{\partial t}
-\nabla\delta\chi^*\nabla\chi-\frac{const}{4}\,\delta\chi^*\chi+c.c.\bigg){\rm d}V{\rm d}t=0,
\ee
where $\delta$ denotes the variation of the corresponding quantity.

Now, we perform integration by parts with respect to time in the first
term and with respect to the spatial coordinates in the second one 
and assume that the variations of $\chi$ equal zero at the 
borders of the integration region
\be
\label{var1}
\delta\chi|_{t=0}^{\infty}=0,
\ee
\be
\label{var2}
\delta\chi|_{x_k=-\infty}^{x_k=\infty}=0,\quad k=1,2,3.
\ee
Then we get the equation
\be
\label{Kvv3}
\int_0^{\infty}\int\delta\chi^*\bigg(\Delta
-\frac{1}{c^2}\frac{\partial^2}{\partial t^2}
-\frac{const}{4}\bigg)\chi\,{\rm d}V{\rm d}t+c.c.=0
\ee
that has to be obeyed for arbitrary variations $\delta\chi$
and $\delta\chi^*$. 
It yields the equation for the probability amplitude $\chi$ in the form
\be
\label{Klein}
\bigg(\Delta-\frac{1}{c^2}\frac{\partial^2}{\partial t^2}
-\frac{const}{4}\bigg)\chi=0
\ee
and its complex conjugate.
Using Eq. (\ref{chi}) the last equation becomes
\be
\label{Kvv4}
(\Delta\psi)\eta
-\frac{1}{c^2}\bigg(\frac{\partial^2\psi}{\partial t^2}\eta
+2\frac{\partial\psi}{\partial t}\frac{{\rm d}\eta}{{\rm d}t}
+\psi\frac{{\rm d}^2\eta}{{\rm d}t^2}\bigg)
-\frac{const}{4}\psi\eta=0.
\ee
For a particle with the infinite life time, we assume
that the time derivatives of the function $\eta$ go to zero, 
${\rm d}\eta/{\rm d}t\rightarrow 0$ 
and ${\rm d^2}\eta/{\rm d}t^2\rightarrow 0$.
Then, using the probability amplitude (\ref{free}) 
and the constant $const$ in form (\ref{const}) we obtain
the usual Klein-Gordon equation for a free particle
\be
\label{KleinG}
\bigg(\Delta-\frac{1}{c^2}\frac{\partial^2}{\partial t^2}
-\frac{m_0^2c^2}{\hbar^2}\bigg)\psi=0.
\ee
At the same time, the normalization condition (\ref{norm})
has to be replaced by Eq. (\ref{norm1}).

The non-relativistic time Schr\"odinger equation for a free particle
\be
\label{Schrod}
i\hbar\frac{\partial\varphi}{\partial t}=-\frac{\hbar^2}{2m_0}\Delta\varphi
\ee
can be obtained from the Klein-Gordon equation (\ref{KleinG}) by using the transformation
\be
\label{Spsi}
\psi={\rm e}^{m_0 c^2 t/(i\hbar)}\varphi,
\ee
where $\varphi$ is the probability amplitude appearing in the Schr\"odinger
equation.
This transition is well-known and will not be discussed here (see
e.g. \cite{Davydov,Shankar}).
For different derivation of the Klein-Gordon equation and 
time Schr\"odinger equation, see \cite{Reginatto}.

The Dirac equation can be derived by taking the probability
amplitude $\chi$ in Eq. (\ref{Kvv1}) in form of a column vector with four components
\be
\label{KvvD0}
\int_0^{\infty}\int
\bigg(\frac{1}{c^2}\frac{\partial\chi^+}{\partial t}\frac{\partial\chi}{\partial t}
-\sum_{k=1}^3\frac{\partial\chi^+}{\partial x_k}\frac{\partial\chi}{\partial x_k}
-\frac{const}{4}\chi^+\chi\bigg){\rm d}V{\rm d}t=0,
\ee
where the cross denotes the hermitian conjugate.
Inserting the $\gamma^{\mu}$ matrices with the well-known properties \cite{Davydov}
into this equation and using Eq. (\ref{const}) we get the integral
\be
\label{KvvD}
K[\chi]=\int_0^{\infty}\int
\bigg[\frac{1}{c^2}
\bigg(\gamma^0\frac{\partial\chi}{\partial t}\bigg)^+
\bigg(\gamma^0\frac{\partial\chi}{\partial t}\bigg)-
\sum_{k=1}^3
\bigg(\gamma^k\frac{\partial\chi}{\partial x_k}\bigg)^+
\bigg(\gamma^k\frac{\partial\chi}{\partial x_k}\bigg)
-\frac{m_0^2 c^2}{\hbar^2}\chi^+\chi\bigg]{\rm d}V{\rm d}t=0
\ee
analogous to the integral $J[\chi]$.
Then, using properties of the $\gamma^{\mu}$ matrices,
performing the integration by parts and using boundary conditions
analogous to Eqs. (\ref{var1}) and (\ref{var2}) 
we get
(see also \cite{Frieden})
\be
\label{Dvv3}
K[\chi]=\int_0^{\infty}
\int\bigg(\frac{\gamma^0}{c}\frac{\partial\chi}{\partial t}
-\sum_{k=1}^3 \gamma^k\frac{\partial\chi}{\partial x_k}
-i\frac{m_0 c}{\hbar}\chi\bigg)^+
\bigg(\frac{\gamma^0}{c}\frac{\partial\chi}{\partial t}
+\sum_{k=1}^3\gamma^k\frac{\partial\chi}{\partial x_k} 
+i\frac{m_0 c}{\hbar}\chi\bigg){\rm d} V{\rm d}t=0.
\ee
The operator in the first parentheses is the hermitian
conjugate of that in the second ones.
Equation (\ref{Dvv3}) can be obeyed by assuming that the
expression in the first or second parentheses equals zero. 
After the substitution $\chi=\psi\eta$ and assumption 
${\rm d}\eta/{\rm d}t\rightarrow 0$
the latter condition yields the Dirac equation in the form
\be
\label{Diraceq}
\frac{\gamma^0}{c}\frac{\partial\psi}{\partial t}
+\sum_{k=1}^3\gamma^k\frac{\partial\psi}{\partial x_k}+
\frac{im_0c}{\hbar}\psi=0.
\ee

We have seen that requirement of the relativistic invariance
of the linear combination of the time and spatial Fisher informations yields 
the basic equations of motion of quantum mechanics.
The scalar and vector potentials $U$ and ${\bf A}$ can be included into the equations 
of motion by means of the usual rules
$i\hbar(\partial/\partial t)\rightarrow i\hbar(\partial/\partial t)-qU$
and 
$-i\hbar\nabla\rightarrow -i\hbar\nabla-q{\bf A}$ discussed above.

\section{Classical mechanics}

\label{classical}

To derive the Hamilton-Jacobi equation of classical mechanics we proceed as follows. 
The probability amplitude $\varphi$ appearing in the Schr\"odinger
equation can be taken in the form analogous to Eq. (\ref{psi12})
\be
\label{rvarphi}
\varphi({\bf r},t)={\rm e}^{i(S_1+iS_2)/\hbar}={\rm e}^{iS_1/\hbar}{\rm e}^{-S_2/\hbar},
\ee
where the phase of the probability amplitude is now expressed in the units $\hbar$
and the probability amplitude is normalized by means of the usual
condition $\int |\varphi|^2{\rm d}V=1$. 
The time Schr\"odinger equation with the potential energy $qU$
equals
\be
\label{SchrU}
i\hbar\frac{\partial\varphi}{\partial t}= 
-\frac{\hbar^2}{2m_0}\Delta\varphi+qU\varphi.
\ee
Multiplying this equation by $\varphi^*$ and integrating over the whole
space we get after simple integration by parts
\be
i\hbar\int\varphi^*\frac{\partial\varphi}{\partial t}{\rm d}V=
\frac{1}{2m_0}\int|-i\hbar\nabla\varphi|^2{\rm d}V+\int qU|\varphi|^2{\rm d}V.
\ee
Substituting Eq. (\ref{rvarphi}) into the last equation yields
\be
\label{Schr2}
\int\frac{\partial S_1}{\partial t}{\rm e}^{-2S_2/\hbar}{\rm d}V+
i\int\frac{\partial S_2}{\partial t}{\rm e}^{-2S_2/\hbar}{\rm d}V+
\frac{1}{2m_0}\int(\nabla S_1)^2{\rm e}^{-2S_2/\hbar}{\rm d}V+
\ee
\bd
+\frac{1}{2m_0}\int(\nabla S_2)^2{\rm e}^{-2S_2/\hbar}{\rm d}V
+\int qU{\rm e}^{-2S_2/\hbar}{\rm d}V=0.
\ed
Due to the normalization condition 
$\int |\varphi|^2 {\rm d}V=\int \exp(-2S_2/\hbar) {\rm d}V=1$, 
the second integral in this equation equals zero
and all imaginary terms disappear from this equation.

Now we assume that the probability density
\be
\label{rrho}
|\varphi|^2={\rm e}^{-2S_2/\hbar}
\ee 
has very small values everywhere except for the close vicinity of the point 
$\langle{\bf r}\rangle=\langle{\bf r}\rangle(t)$, 
where it achieves its maximum and the first derivatives of $S_2$ 
at this point equal zero  
\be
\label{rder}
\left.\nabla S_2\right|_{{\bf r}=
\langle{\bf r}\rangle} =0.
\ee
In such a case, the probability density can be replaced by
the $\delta$-function
\be
\label{rdelta}
|\varphi|^2=
\delta({\bf r}-\langle {\bf r}\rangle)
\ee
and the probabilistic character of the theory disappears.
Therefore, the function $S_2$ describing the form of 
the probability distribution $\rho$ does not appear in classical mechanics.

Then, straightforward use of Eqs. (\ref{Schr2})-(\ref{rdelta}) leads to the 
Hamilton-Jacobi equation for the function $S_1$ 
in the variable $\langle {\bf r}\rangle$
\be
\label{HJ0}
\frac{\partial S_1(\langle {\bf r}\rangle,t)}{\partial t}+
\frac{(\nabla S_1(\langle {\bf r}\rangle,t))^2}{2m_0}
+qU(\langle {\bf r}\rangle,t)=0.
\ee
In classical mechanics, the variable $\langle {\bf r}\rangle$ is
usually replaced by the classical coordinate ${\bf r}$ and the
function $S_1$ is called the action $S$
\be
\label{HJ}
\frac{\partial S({\bf r},t)}{\partial t}+
\frac{(\nabla S({\bf r},t))^2}{2m_0}
+qU({\bf r},t)=0.
\ee

We note that Eq. (\ref{rdelta}) corresponds to the limit 
$S_2\gg \hbar$ or $\hbar\rightarrow 0+$ in Eq. (\ref{rrho}).
Therefore, the function $S_1$ in Eq. (\ref{HJ0}) is in fact the first term 
of the expansion of $S_1$ into the power series in $\hbar$
\be
S_1=S_1|_{\hbar=0}+\cdots.
\ee
In this limit, the commutation and uncertainty relations
disappear from the theory.

We have seen that the Hamilton-Jacobi equation follows
from the probabilistic description of results of measurements
in the limit of the $\delta$-like probability densities $\rho$ and 
the non-relativistic approximation.
As usual, the vector potential ${\bf A}$ 
can be included into the theory by means of the rule 
$\nabla S\rightarrow \nabla S-q{\bf A}$
following from the rule
$-i\hbar\nabla\rightarrow -i\hbar\nabla-q{\bf A}$ discussed 
in sections \ref{vector} and \ref{scalar}.
The Hamilton-Jacobi equation can be also the starting point
for deriving the Hamilton variational principle of classical
mechanics.

\section{Many particle systems}

\label{many}

The starting point of discussion of the $N$ particle system is 
definition analogous to Eq. (\ref{meanx})  
\be
\label{meanxN}
\langle {\bf r}_j \rangle=
\int {\bf r}_j\rho({\bf r}_1,\ldots,{\bf r}_N,t){\rm d} V_1\ldots{\rm d}V_N,
\quad j=1,\ldots,N,
\ee
where $\rho$ is the many particle probability density
and ${\bf r}_j$ are the coordinates of the $j$-th particle.
Then, discussion can be performed analogously to that given
above and the probability amplitude, uncertainty and commutation
relations, momentum operators and density currents for all
particles can be introduced. 
The scalar and vector potentials $U({\bf r}_1,\ldots,{\bf r}_N,t)$ and
${\bf A}({\bf r}_1,\ldots,{\bf r}_N,t)$
and antiparticles can be also discussed.  

The Schr\"odinger equation for $N$ free particles can be found from 
the many particle generalization of the relativistic invariant (\ref{Kvv1}) 
\be
\label{KvvN}
\int_0^{\infty}\int
\bigg(\frac{1}{c^2}\bigg|\frac{\partial\chi}{\partial t}\bigg|^2
-\sum_{j=1}^N|\nabla_j\chi|^2
-\sum_{j=1}^N\frac{m_j^2c^2}{\hbar^2}|\chi|^2\bigg){\rm d}V_1\ldots{\rm d}V_N{\rm d}t=0,
\ee
where $\chi({\bf r}_1,\ldots,{\bf r}_N,t)$ is the $N$ particle
probability amplitude
and $m_j$ denotes the rest mass of the $j$-th particle.
Using similar approach as in section \ref{equations}, we can then obtain the 
Schr\"odinger equation for $N$ free particles
\be
\label{SchrodN}
i\hbar\frac{\partial\psi}{\partial t}=-\sum_{j=1}^N\frac{\hbar^2}{2m_j}\Delta_j\psi
\ee
and the Hamilton-Jacobi equation
\be
\label{rHJclN}
\frac{\partial S}{\partial t}+\sum_{j=1}^N\frac{(\nabla_j S)^2}{2m_j}=0.
\ee

It seen that the probabilistic description of results of measurements 
and its relativistic invariance yield also the basic
mathematical structure of the many particle quantum mechanics.

\section{Postulates of quantum mechanics}

\label{postulates}

In this section, we will show that the
basic formulation of the postulates of quantum mechanics
can be obtained from the discussion given in the preceding sections.

It follows from the above discussion that the probability
density $\rho({\bf r},t)$ and the probability density current
$j_k({\bf r},t)$ can be in a mathematically simple and straightforward way
represented by the probability amplitude $\psi({\bf r},t)$.
Using the probability amplitude $\psi({\bf r},t)$, it is possible
to calculate the mean values of the operators representing 
the coordinates $x_k$ and momentum $p_k$, $k=1,2,3$, and
further physical quantities that can be expressed in terms of the coordinates
and momentum (the kinetic energy, total energy, etc.). 
In this sense, the probability amplitude $\psi({\bf r},t)$ represents
the state of the system.
It agrees with the usual {\em first postulate} of quantum mechanics
formulated in the coordinate representation.

According to some interpretations of quantum mechanics, the probability
amplitude $\psi$ represents only the subjective knowledge of the observer. 
However, since $\psi$ carries the information on $\rho$ and $j_k$,
it contains the objective physical information on the system. 
Certain analogy with this situation can be found in statistical physics, 
where the statistical distribution functions describe physical reality
in a different probabilistic sense.
For this reason, we do not agree with such subjective interpretations.
       
Mean values of the coordinates and momentum can be calculated from
the formulas (compare with Eqs. (\ref{a}) and (\ref{b}))
\be
\langle x_k\rangle=\int\psi^* \hat{x}_k \psi {\rm d}V
\ee
and
\be
\langle p_k\rangle=\int\psi^*\hat{p}_k\psi{\rm d}V.
\ee
Here, the coordinate and momentum operators equal
$\hat{x}_k=x_k$ and $\hat{p}_k=-i\hbar(\partial/\partial x_k)$,
where $\hbar$ depends on the units used in measurement.
To obtain the Hamilton operator $\hat{H}=\hat{T}+\hat{V}$
appearing in the Schr\"odinger equation, the operators
$\hat{x}_k$ and $\hat{p}_k$ can be formally substituted
for the classical quantities $x_k$ and $p_k$ 
in the classical Hamilton function 
\be
H=T+V=\frac{1}{2m_0}\sum_{k=1}^3 
\left(\frac{\partial S}{\partial x_k}\right)^2+V=
\frac{1}{2m_0}\sum_{k=1}^3 (p_k)^2+V
\ee 
appearing in the Hamilton-Jacobi equation (\ref{HJ}).
It is obvious that similar rule applies also for other physical quantities
depending on $x_k$ and $p_k$.
This rule agrees with the {\em second postulate} of quantum mechanics.   

For the sake of simplicity, the following discussion will be made 
in one dimension only.
In our original $x$--representation, we assumed that we 
use the experimental apparatus that can measure
the coordinate and the probability density of getting the coordinate $x$
in measurement equals $\rho(x)=|\psi(x)|^2$ (see Eq. (\ref{meanx})). 
Let us assume now that we perform measurement with another 
experimental apparatus that measures the momentum and 
the probability of getting the value $p$ of the momentum
equals $|\varphi(p)|^2$.
We assume also that the corresponding probability amplitudes
$\psi$ and $\varphi$ are related by the unitary transformation
$U$ preserving the norm of the probability distributions 
\be
\psi=U\varphi.
\ee
Then, the mean value of the coordinate $x$
equals 
\be
\langle x\rangle=\int\psi^* x\psi{\rm d}x=
\int \varphi^* (U^{-1}x U)\varphi {\rm d}p,
\ee 
where the integration in the second integral is performed over $p$.
It means that the coordinate operator in this momentum
or $p$--representation is not diagonal $\hat{x}=U^{-1}x U$.
Assuming that the system that will be measured is prepared 
in the state described by the probability amplitude $\varphi$,
the probability of getting the value $x$ in the following measurement
of the coordinate $x$ equals
\be
\langle x\rangle=\int x|\psi|^2{\rm d}x,
\ee
where 
\be
|\psi|^2=|U\varphi|^2.
\ee
This result agrees with the {\em third postulate} of quantum mechanics,
according to which the probability of getting the value $x$
in measurement of the coordinate on the system in the state described by 
the wave function $\varphi$ can be expressed as $|\langle x|\varphi\rangle|^2$.

We note that the values of the coordinate $x$ appearing in the definition of the mean value 
$\langle x\rangle$, Eq. (\ref{meanx}), or the values of the momentum $p$ in measurement of
the momentum are real quantities and the corresponding operators $\hat{x}$
or $\hat{p}$ are diagonal in their own $x$-- or $p$--representation.
After transition to another representation by means of 
the unitary representation, these operators are not diagonal,
however, they remain hermitian.
Also this conclusion agrees with the rules of quantum mechanics.
 
After performed measurement of the coordinate $x$ giving one concrete
value $x'$, the state of the system can be described by 
the probability amplitude $\delta(x-x')$ (the eigenfunction
of the operator $\hat{x}=x$).
This probability amplitude replaces the original one 
describing the state of the system immediately before the measurement and its 
time evolution till the time of the following measurement is
given by the corresponding equation of motion. 
This change of the probability amplitude resulting from
the performed measurement is
known as the {\em reduction} or {\em collapse} of the wave function.        

Requirement of the relativistic invariance of the generalized
space-time Fisher information yields the usual 
{\em fourth postulate} of quantum mechanics postulating
the Schr\"odinger equation (see section \ref{equations}).
 
We have seen that the rules discussed above can be obtained from 
the probabilistic description of measurement of the coordinates and time and 
the relativistic invariance of the theory.
More general formulation of the postulates of quantum mechanics can be
obtained by generalizing
the above discussion to further physical quantities.

\section{Conclusions}

In \cite{Lande}, Land\'e tried to at least partly illuminate 
the old problem of Einstein {\em why the world is a quantum world}
\cite{Einstein}.  
Land\'e started his discussion of the quantum world with three
postulates: 
(a) Symmetry of the probabilities of the transition from the state $\alpha$
to the state $\beta$, $P_{\alpha\beta}=P_{\beta\alpha}$.
(b) Correspondence between the actual microscopic and 
the ordinary probability law.
(c) Covariance of dynamics requiring that only differences
of physical quantities like the energy appear in observable quantities.   
Replacing the standard postulates by the postulates (a)-(c),
he was able to obtain the momentum 
operator and commutation relations. 
Using the coordinate and the momentum operator, 
the Schr\"odinger equation can be obtained in the usual way
from the classical mechanics.  

In this paper, we have made an attempt to avoid this or another set of similar postulates 
and clarify the basic structure of quantum mechanics by using even more general approach.

We started by assuming that the probabilistic results of measurement of the space coordinate
$x$ are given by the probability density $\rho({\bf r},t)$.
Then, using integration by parts and the Schwarz inequality, we derived 
relations (\ref{Schwarz}) containing an arbitrary complex function $\psi({\bf r},t)$.
In general, quantities $\int x^n|\psi|^2{\rm d}V$ appearing in these relations
are different from the moments $\langle x^n\rangle=\int x^n\rho\,{\rm d}V$
characterizing results of measurements.
Requiring that relations (\ref{Schwarz}) contain physically relevant
moments $\langle x^n\rangle$ and $\int|\psi|^2{\rm d}V=1$, we concluded that the most simple solution
of these requirements has the form of the Born rule $\rho=|\psi|^2$ (see section \ref{Bornrule}).
At this point, the physical meaning of the phase $s_1$ of the probability amplitude 
$\psi=|\psi|\exp(is_1)=\exp(-s_2)\exp(is_1)$ is not obvious.     
 
In section \ref{commutation}, we used the normalization condition $\int\rho\,{\rm d}V=1$ and
the Born rule $\rho=|\psi|^2$ and applying the integration by parts we obtained the commutation
relation (\ref{apx2}). 
This commutation relation depends on $s_2$ only and disappears if the 
theory has not probabilistic character, as for example in case of classical mechanics.
Similar commutation relation should appear in any probabilistic theory
analogous to that described in section \ref{Bornrule}.

In the following section \ref{uncertainty}, we started with the uncertainty relation
(\ref{Schwarz}) obtained in section \ref{Bornrule} and derived a few 
related uncertainty relations including the Heisenberg uncertainty
relation in form (\ref{cSchwarz2}) that does not contain the Planck constant
$\hbar$.
We found also the uncertainty relation (\ref{Heiss2}) that does not depend on $s_1$ 
and its left hand side is less than or equal to that in relation (\ref{cSchwarz2}).
Analogously to the commutation relations, the uncertainty relations
should appear in any probabilistic theory analogous to that discussed
in this paper.
In the limit case in which the probabilistic character of the theory disappears,
the commutation and uncertainty relations disappear, too.

The commutation relation (\ref{apx2}) remains valid if the operator 
$-i(\partial/\partial x)$ is replaced by the operator 
$-i(\partial/\partial x)-f_x({\bf r},t)$, where $f_x$ is an arbitrary
real function. 
It makes possible to generalize the theory and include for example 
the vector potential describing the interaction with the
electromagnetic field (section \ref{vector}).

Physical meaning of the phase $s_1$ was discussed in section \ref{probability}.  
To describe motion of particles in space, it is necessary to introduce not
only the probability distribution $\rho$ but also the corresponding 
probability density current in the form ${\bf j}=\rho\,{\bf v}$.
Here, the "velocity" can be written in a general form ${\bf v}=\hbar\nabla s_1/m_0$,
where $s_1$ is the phase of the probability amplitude, $m_0$ is the rest mass
of the particle and $\hbar$ is a constant that has to be determined experimentally.
Thus, the probability amplitude $\psi=\exp[i(s_1+is_2)]$ carries
information on the form of the probability density current ${\bf j}$ 
contained in $s_1$ and the probability distribution $\rho$ contained in $s_2=-(1/2)\ln\rho$.
The operator $-i(\partial/\partial x)$ appearing in the expression
(\ref{j2}) for the probability density current is needed
for describing the motion of particles in space.
    
The Fisher information characterizing the smoothness of the probability
distribution $\rho$ appears already in Eq. (\ref{Schwarz}). 
Its generalization including not only $s_2$ but also $s_1$ was discussed
in section \ref{Fisher}.
Except for a factor, this generalization of the Fisher information 
agrees with the kinetic energy known form quantum mechanics and
shows its importance in physics.
The momentum operator was discussed in section \ref{Fisher}, too.

To discuss time in section \ref{time}, 
we assumed that the investigated system has a finite
life time and the probability density can be normalized by
the condition $\int_{t=0}^{\infty}\int\rho({\bf r},t){\rm d}V{\rm d}t=1$.
Then, using similar approach as for the coordinate $x$, 
we introduced the time component of the probability density current $j_t$,
the operator $i(\partial/\partial t)$ and derived 
the time--energy uncertainty relation (\ref{tuuvv2}).
Different roles of time and the spatial coordinates were also discussed.

Analogously to the vector potential, we discussed also the
scalar potential.
The unidirectional character of time makes possible to understand also
the existence of antiparticles (section \ref{scalar}).  

Equations of motion are derived from the requirement that a linear
combination of the space and time Fisher informations is a relativistic
invariant (section \ref{equations}).
It leads not only to the Klein-Gordon, Schr\"odinger and Dirac equation,
but it helps also to understand some assumptions made in the special
theory relativity.

Transition from quantum to classical mechanics is well-known.
Therefore, only brief discussion of the transition from the Schr\"odinger
equation to the Hamilton-Jacobi equation of classical mechanics was 
made in section \ref{classical}.
 
Generalization to many-particle systems is straightforward 
and was made in section \ref{many}.

The standard postulates of quantum mechanics and their relation to
our approach were discussed in section \ref{postulates}.

Discussion in the first part of this paper can be applied 
not only to the spatial coordinates $x$, $y$ and $z$, but also to some other
physical quantities.
However, it does not apply for the equations of motion that are derived
from the requirement of the relativistic invariance of the space-time
Fisher information $I''_t/c^2-(I''_x+I''_y+I''_z)$ depending on $x$, $y$, $z$ and $t$.    

As we mentioned in section \ref{Bornrule}, 
our way of writing the probability amplitude
in Eq. (\ref{psieik}) is very similar to that used by Bohm
\cite{Bohm1,Bohm2}.
However, our main results and their interpretation are different from those of Bohm.
Since there is no experimental evidence for a precisely defined 
position of particles assumed in \cite{Bohm1,Bohm2}, we do not make such
assumption here.
Also, we do not introduce any additional quantum potential like Bohm
since such a potential would violate the commutation 
relations that are general consequence of the probabilistic
description of measurements.
In contrast to \cite{Bohm1,Bohm2}, where the validity of 
the Schr\"odinger equation is postulated 
we require here the relativistic invariance of the generalized space-time 
Fisher information leading to the relativistic equations of motion.
We note also that no hidden variables are needed in our approach. 
Therefore, our understanding of quantum mechanics is close to the 
standard or Copenhagen interpretation and differs from that of Bohm.

We note also that formulating the theory in terms of the probability
amplitude $\psi$ instead the original probability density $\rho$ and the probability
density current ${\bf j}$ has a very significant advantage, namely, 
the equations of motion are linear in $\psi$.
Thus, approach based on the probability amplitude leads to 
mathematically simple theory based on the linear vector spaces
in which the superposition principle for $\psi$ is valid. 
If another quantities like $\rho$ and ${\bf j}$ or $s_1$ and $s_2$
were used for representing the state of the system, 
this physically important property valid for $\psi$ would be lost.
 
Another question is whether the relation between the probability amplitude
$\psi$ and the probability density $\rho$ 
could be different from that given by the Born rule (\ref{Born}) or (\ref{psi}).
For example, we can replace Eq. (\ref{psi}) by the equation     
\be
\xi=\rho^{1/4}\, {\rm e}^{is_1/2}.
\ee
Then, we obtain the normalization condition
\be
\int|\xi|^4{\rm d}V=1
\ee
and the mean square coordinate
\be
\langle x^2\rangle=\int x^2|\xi|^4{\rm d}V.
\ee
Analogous formulae can be obtained also
for other physical quantities by replacing 
the usual probability amplitude $\psi$
by $\xi^2$.
Resulting formulae that can be obtained from  
equations in sections \ref{Bornrule}-\ref{postulates} are more complicated 
and the equations of motion are not linear in
$\xi$.
Therefore, the most simple theory with the probability amplitudes 
creating the linear vector spaces is obtained only
for the usual relation between the probability
density and probability amplitude in form of the Born rule.  

Unperformed experiments have no results. 
Therefore, it follows from our discussion that quantum mechanics 
does not speak explicitly of events in the measured system, but only of results
of measurements, implying the existence of external measuring apparatuses
that has been supposed above.

Of course, one can argue that some steps of our discussion 
were motivated by our knowledge of quantum mechanics.
For this reason, the most important steps were discussed in detail and 
their relation to standard quantum mechanics was clarified.
The resulting way of obtaining quantum mechanics is not unique. 
However, our discussion shows that quantum mechanics is, in the sense of
the Occam's razor, apparently the
most simple and straightforward way of describing
the probabilistic nature of certain class of physical phenomena.
It shows also that the main ideas of quantum mechanics are 
understandable and physically and mathematically transparent. 

Now we can return to the Einstein's question 
{\em why the world is a quantum world}.
Taking into consideration that 
the interaction of the measured system with the measuring apparatus
and the rest of the world cannot be in general neglected, measuring apparatuses are
not described in detail but on the macroscopic level only, 
real physical detectors have limited resolution and efficiency and 
the experimental control of the initial conditions
is limited, the probabilistic description of results of measurements
seems to be unavoidable.
As shown in this paper, the probabilistic description of results of measurements then leads, 
together with the relativistic invariance of the theory, to quantum mechanics.
The deterministic description of the world in the sense of
classical mechanics is possible only in special cases when
the probabilistic character of measurements can be neglected.

Since the information theoretical approach used in this
and some other papers 
(see e.g. \cite{Bohr,Hardy,Summhammer,Parwani,Frieden,Brukner,Zeilinger3,Hall1,Fivel,Reginatto,Frohner})
makes possible to obtain the most significant parts 
of the mathematical formalism of quantum mechanics from
the probabilistic description of results of measurements, 
we believe that it is general and physically correct 
starting point to understanding this field.
It shows that the roots of
quantum mechanics are deeply related to the probability theory 
and helps to understand quantum theory as correctly
formulated probabilistic theory
that can describe certain class of physical phenomena        
at different levels of accuracy from the most simple models to 
very complex ones.

\acknowledgments{
This work was supported by
the MSMT grant No. 0021620835 of the Czech Republic.}

\end{document}